# NEW HORIZONTAL FRUSTUM OPTICAL WAVEGUIDE FABRICATION USING UV PROXIMITY PRINTING

*Tsung-Hung Lin[1], Hsiharng Yang[2], Ruey Fang Shyu[3], and Ching-Kong Chao[1]*

[1] Department of Mechanical Engineering, National Taiwan University of Science and Technology, Taipei, Taiwan 105
[2] Institute of Precision Engineering, National Chung Hsing University, Taichung, Taiwan 402
3 Department of Mechanical Manufacturing Engineering, National Formosa University, Taiwan 632

**ABSTRACT**

This paper presents a novel method to fabricate the horizontal frustum structure as a planar optical waveguide by using the proximity printing technique. A horizontal frustum optical waveguide with a both lateral and vertical taper structure was produced. The orthogonal and inclined masks with the diffraction effect were employed in lithography process. This method can precisely control each horizontal frustum optical waveguide geometric profile in the fabrication process. The horizontal frustum optical waveguide and its array with the same inclined angle were generated. The beam propagation simulation software (BPM_CAD) was used to modeling the optical performance. The simulation results reveal that the mode profile matched into horizontal frustum optical waveguide and fiber from the laser diode. The optical loss of horizontal hemi-frustum structure of optical waveguides was less than 0.2dB. The horizontal hemi-frustum waveguide will be used for fiber coupling on boards for further optical communication systems.

## 1. INTRODUCTION

Integrated Optical Circuits (IOCs) have been under development in many laboratories and companies for over three decades. New access technologies such as Gigabit Ethernet, 10 Gigabit Ethernet, and passive optical access systems are investigated [1]. Optical Broadband Access Technologies (OBAT) with the current dominant broadband access technology of Hybrid-Fiber-Coax (HFC) systems is very interested. In recent years, an optical device integrated with a spot-size converter has been paid much attention for its direct coupling to an optical fiber without a micro-lens [2], tapered fiber [3] or lens fibers [4]. The laser diode generally has a small field radius in order to minimize the pumping current, and also has an elliptically shaped mode profile. An optical fiber has lager dimensions and is circularly symmetric with a mode radius of about 4.5 µm. The main problems associated coupling light from a semiconductor laser diode to an optical fiber lie in the mismatch between the mode profiles of the laser and the fiber, as well as establishing the alignment between them.

A laser diode integrated with a spot-size converter is much more attractive for low-cost packaging due to its large spot-size, which is well matched to that of a single-mode fiber [5]. Three main classes of spot-size converter have been developed to expand the optical mode. The first is the vertical spot-size converter in which the waveguide thickness is decreased along the output direction. The second is laterally spot-size converter in which the waveguide width is decreased. The third is the lateral and the vertical dimensions of the guiding layer are changed. Such a combination allows us to easily control the beam divergence at the output facet [6]. These kinds of tapered waveguides have been proposed and demonstrated to improve the optical coupling among optoelectronic devices [7, 8, 9].

The taper waveguide exist much fabrication methods. The laterally taper can be achieved by standard photolithography followed by wet chemical etching, reactive ion etching (RIE) or reactive ion beam etching (RIBE) [10]. To fabricate vertical taper methods are including shadow mask techniques [11, 12], selective growth [13, 14, 15], sulfuric acid dip-etching [16], and diffusion-limited etching with selective area epitaxy [17]. The other type uses both lateral and vertical structuring, which is generally harder to fabricate since most current processes are planar. It is very difficult to achieve tapers in the vertical direction, but it has been done using such techniques as utilizes the selective area growth to add reactive ion etching or Stepped etching to add $Cl_2$ chemical dry etching and regrowth [18, 19]. However, their processes are limited to semiconductor materials. Using a planar technology, such as spin coating of polymers, it is not easy to make structures which have a physical vertical taper shape. It can be done using laser ablation or RIE, but with an increased fabrication cost. The two guiding layers are then patterned using different masks. This is very hard to do using semiconductors, since it involves two mask, etch and regrowth steps and various resists and etches have to be applied to and cleaned from the intermediate surfaces [20]. In this study, the fabrication of horizontal hemi-frustum optical waveguides is proposed here to provide a novel method. A horizontal hemi-frustum optical waveguide with a lateral and vertical taper structure was produced.

## 2. LITHOGRAPHY CHARACTERISTICS

Lithographic exposure has main three printing mode that contact, proximity, and projection. The contact printer involves putting a photomask in direct contact with the wafer. The contact printing with zero gap distance ideally offers best resolution shown in Figure 1(a). The exposure operations using proximity mode is in the near field or Fresnel diffraction regime. The pattern resulting from the light passing through the mask directly impacts onto the photoresist surface because there is no lens between the photoresist and mask. The created aerial image therefore depends on the near field diffraction pattern. Because of the diffraction effects, the light bends away from the aperture edges and produces partial exposure outside the aperture edges.

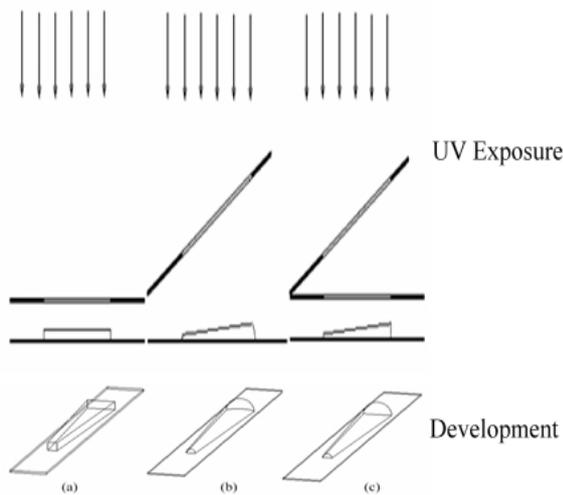

Figure1. Illustration of the microstructure was using the three exposure mode after development.

Figure 1(b) shows as the printing gap between the photoresist and longer parallel side of pattern on the mask gradually increased, the height in the waveguide at the lateral sideline became higher. A small printing gap and shorter parallel side of pattern on the mask is lower because the light intensity distribution in the photoresist will no have enough diffraction. The similar horizontal hemi-frustum structure formed after the development process because shorter parallel side and longer parallel side were not flat. The horizontal hemi-frustum structure formed after the development process was using the orthogonal and inclined masks in lithography process shown in Figure 1(c). The light passing through the orthogonal mask directly impacts onto the photoresist surface shorter parallel side and longer parallel side were flat.

## 3. HEMI-FRUSTUM OPTICAL WAVEGUIDE FABRICATION

Desired patterns are transferred from the designed mask in the lithographic process. In this experiment, a plastic mask was fabricated using laser writing onto a PET (Polyethylene terephthalate) used for PCBs (print circuit boards). The layout pattern on the orthogonal and inclined plastic masks is illustrated in Figure 2. The altitude of trapezoid was L=1000μm on the inclined masks shown in Figure 2(a). The tilted angle $\theta$ was between orthogonal and inclined masks. The altitude of trapezoid was $L\cos\theta$ μm on the orthogonal mask shown in Figure 2(b).

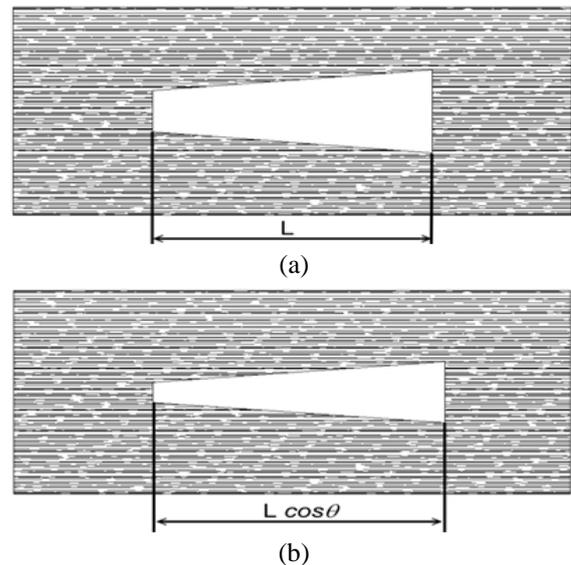

(a)

(b)

Figure 2. Illustration of the pattern design on the mask for Hemi-frustum optical waveguide fabrication; (a) the inclined masks, (b) the orthogonal mask (Unit: μm)

PC (Polycarbonate) sheets 2 mm thick were used as the substrate. To increase the adhesive force between the photoresist and substrate, the PC substrate was coated first with a thin layer of HMDS. The PC substrate was then spun with a layer of negative photoresist (JSR THB-120N) 35 μm thick. The spin condition was 1100 rpm for 20 seconds. Prebaking in a convection oven at 90 °C for 90 seconds is a required procedure. This removes the excess solvent from the photoresist and produces a slightly hardened photoresist surface. The mask was not stuck onto the substrate. The sample was exposed through the plastic mask using a UV mask aligner (EVG620).

This aligner had soft, hard contact or proximity exposure modes with NUV (near ultra-violet) wavelength 350-450nm and lamp power range from 200-500 W. A slice of glass was inserted between the orthogonal and inclined masks to create a gap. Exposure was then conducted for about 16 seconds. The exposure process is shown in Figure 3. The three- dimensional waveguide was completed after exposure and dip into the developer for 90 seconds and cleaning with deionized water. Four tilt angles to place the mask in the lithography process are 5, 8, 10and 15˚.

The Hemi-frustum optical waveguide variations in the samples were observed and the optimal gap range determined. The scanning electron microscopy (SEM) and 3D surface profile were used to measure the characteristics of the resulting waveguide structures.

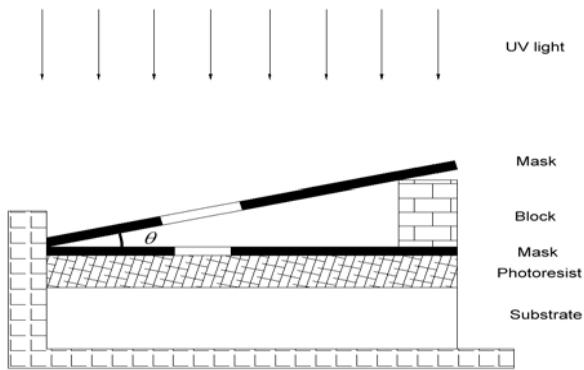

Figure 3. Schematic of the horizontal hemi-frustum optical waveguide fabrication using UV proximity printing.

## 4. RESULTS AND DISCUSSION

According to the experimental results, horizontal hemi-frustum optical waveguides in photoresist were fabricated by using the proximity printing in UV lithography process. The studied photoresist and mask gap sizes ranged from 120 μm to 840 μm. The three horizontal hemi-frustum optical waveguides were classified after development using different gap sizes. A flat top optical waveguides was formed using the printing gap less than 240 μm while the mask tilted 10°. Figure 4 shows the flat top optical waveguide by using SEM observation. Adjusting the printing gap between 240 μm to 840 μm with the mask tilted 10°, horizontal hemi-frustum optical waveguide were formed with 1mm in length as shown in Figure 5. Using the printing gap is larger than 840 μm while the mask tilted 10°, the top view is blurred as observed. This result coincides to previous studied result [21]. Printing gaps ranging from 240 μm to 840 μm using the pattern on the mask can generate optical waveguide in photoresist with vertical taper angle. The horizontal hemi-frustum optical waveguide with vertical heights using titled angles are measured using 3-D surface profiler as shown in Figure 6. The titled angles of 15˚ can generate a higher vertical taper angle. Figure7. Shown the different vertically taper angle resulted from various the mask of tilt angle. The increased mask tilted angle results in a steep variation on the horizontal hemi-frustum optical waveguide with vertical taper angle.

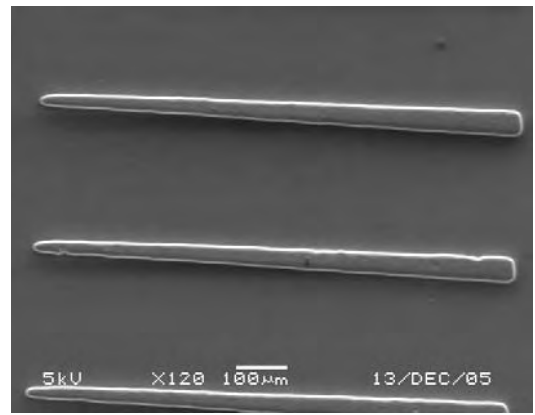

Figure 4. SEM micrograph of horizontal hemi-frustum optical waveguide using 10°tilt angles to place the mask and Printing gaps ranging smaller than 240μm.

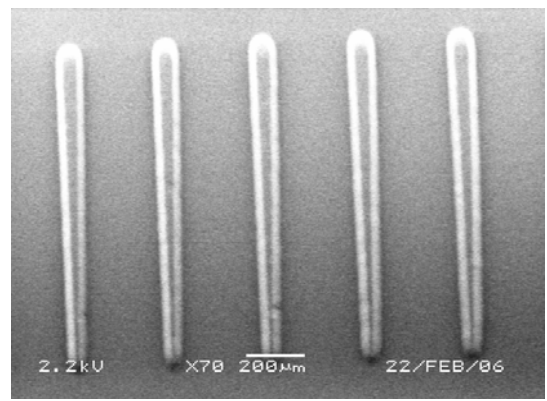

Figure 5. SEM micrograph of horizontal hemi-frustum optical waveguide using 10° tilt angles to place the mask in the lithography process.

TE mode, and TM mode analysis. We launched an infrared light of wavelength 1.55 μm into to horizontal hemi-frustum optical waveguides and single optical fiber form the laser diode in the TE mode. The dimensions of the shorter parallel side length is 3 μm, longer parallel side length is 10 μm, and shorter parallel side high is 2 μm and longer parallel side high is 10 μm the horizontal hemi-frustum optical waveguides. The dimension of the fiber is 9 μm in diameter. Figure 8 shows the optical field distribution in the laser diode, horizontal hemi-frustum optical waveguides and fiber. It is obvious that the mode profiles match fine from laser diode into waveguide and fiber. A minimal of 0.2 dB optical loss can be expected with this type of structures. Different wavelength resulted in the optical loss of horizontal hemi-frustum optical waveguides (shown in Figure 9).

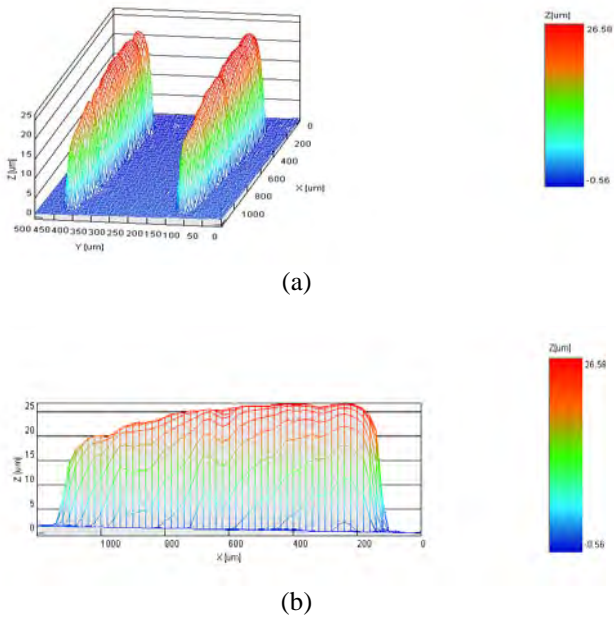

(a)

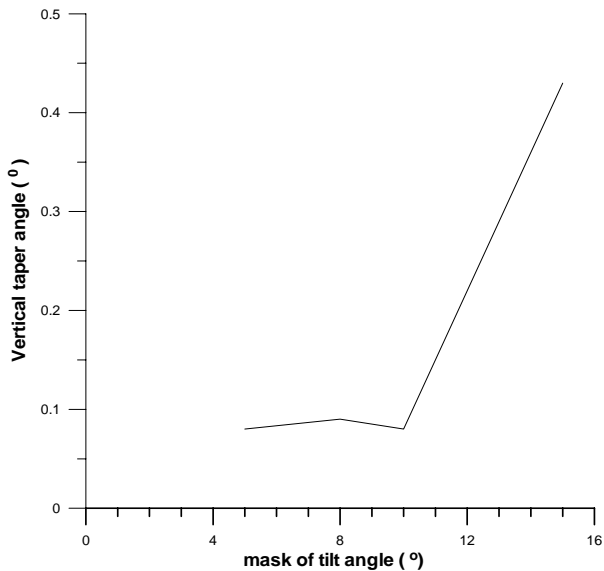

(b)

Figure 6. Experimental results of horizontal hemi-frustum optical waveguide; (a) three-dimensional profile measurement, and (b) cross-sectional profile of the horizontal hemi-frustum optical waveguide.

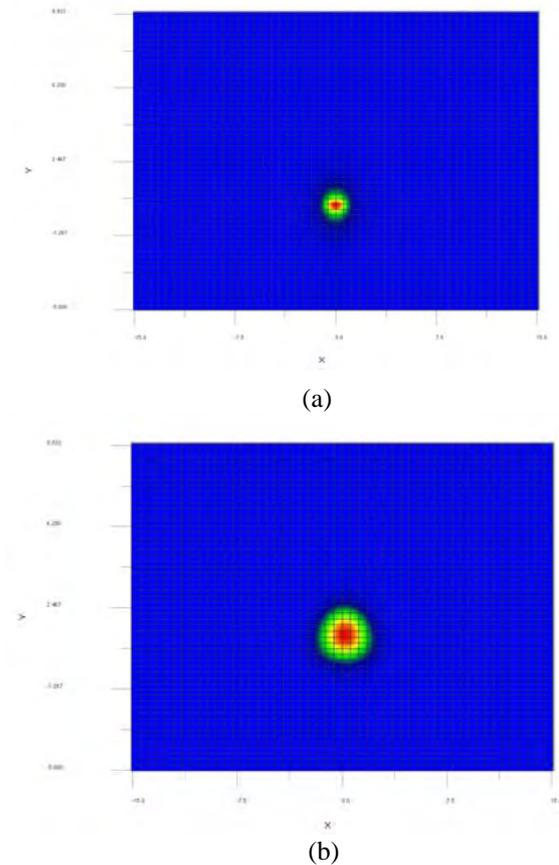

(a)

(b)

Figure 7. Different vertically taper angle resulted from various the mask of tilt angle.

Computer simulation using the beam propagation method (BPM) is implemented to demonstrate the mode-conversion performance of the proposed structures. In the BPM analysis (including mode analysis for initial value), a program that can execute semi-vector analysis is used. This program can execute three kinds of analysis: scalar,

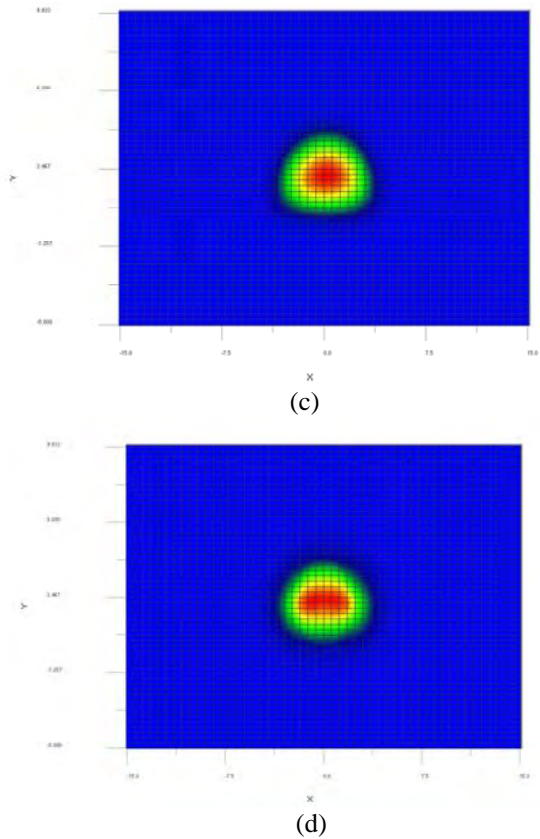

Figure 8. The BPM simulated that to match the mode profiles from laser diode into waveguide and fiber; (a) laser diode, (b) the length of 500 μm of horizontal hemi-frustum optical waveguides, (c) the length of 900 μm of horizontal hemi-frustum optical waveguides, (d) single mode fiber.

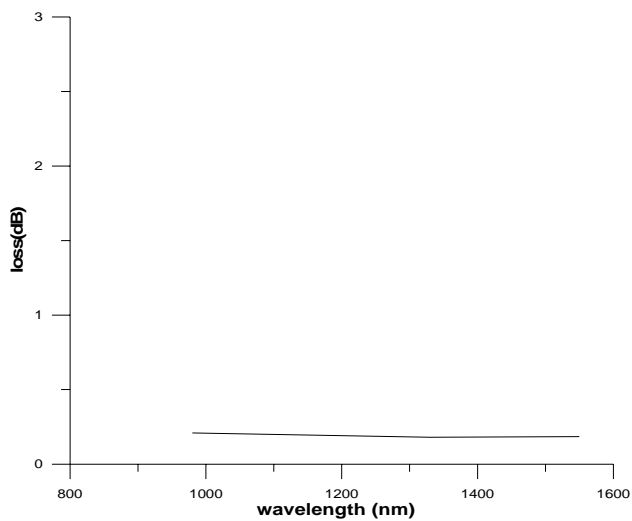

Figure 9. Different wavelength resulted in the optical loss of horizontal hemi-frustum optical waveguides.

## 5. CONCLUSION

The fabrication method of horizontal hemi-frustum optical waveguides for optical communication on board is presented. The horizontal hemi-frustum optical waveguides with the same inclined angle can be controlled using an orthogonal and an inclined mask with a printing gap size. Printing gaps ranging from 240 μm to 720 μm using the pattern on the mask can generate horizontal hemi-frustum optical waveguides in photoresist. High mask tilted angle results in high vertical taper angle on horizontal hemi-frustum optical waveguides. Furthermore, the BPM calculation results reveal that proposed structures have good performances in the transmission efficiency and the fabrication tolerances. The main application of this waveguide can be interchip interconnection for computers and passive components for optical communication systems. This work also offers the novel and simple fabrication method and the potential for reducing the cost due to fiber coupling and packaging of integrated optoelectronic devices.

## 6. ACKNOWLEDGEMENT

This work was supported by the National Science Council (series no. NSC95-2221-E-005-091) of Taiwan, R.O.C. The authors wish to thank Prof. G-K Chang at Georgia Institute of Technology for his research cooperation.